# Durable Enhancement of $MoS_2$ Single-Layer Photoluminescence by Ultraviolet Laser Treatment Under Ambient Conditions

*Mahan Bakhshikhah,[1] Jiří Liška,[1,2] Rahul Kesarwani,[3] Jindřich Mach,[1,2] Ondřej Červinka,[1] Petr Dub,[1,2] Jiří Spousta,[1,2] Jan Přibyl,[4] Jana Kalbáčová Vejpravová,[3] and Tomáš Šikola[1,2*]*

[1]Central European Institute of Technology, Brno University of Technology, 612 00 Brno, Czech Republic

[2]Institute of Physical Engineering, Faculty of Mechanical Engineering, Brno University of Technology, 616 69 Brno, Czech Republic

[3]Department of Condensed Matter Physics, Faculty of Mathematics and Physics, Charles University, 121 16 Prague 2, Czech Republic

[4]Central European Institute of Technology, Masaryk University, 625 00 Brno, Czech Republic

[*]sikola@fme.vutbr.cz

**ABSTRACT**

Single-layer molybdenum disulfide ($MoS_2$) possesses significant potential for nanoscale optoelectronics, but achieving high-intensity, long-term-stable photoluminescence (PL) emission remains a challenge. In this work, we demonstrate a remarkably robust, more than 8-fold maximum enhancement in the PL intensity of exfoliated and CVD-grown single-layer $MoS_2$ via a non-destructive ultraviolet (UV) laser treatment method. This substantial increase in radiative efficiency is accompanied by a trion-to-neutral exciton transition in the PL signal and a corresponding blue shift of the Raman $E^1_{2g}$ and $A_{1g}$ vibrational modes, signaling successful electron depletion (p-doping) and formation of Mo-O bonds, respectively. Furthermore, we demonstrate precise spatial control over PL properties by confining PL treatment exclusively to the UV laser-treated area. Crucially, the enhanced PL performance shows exceptional longevity; the CVD sample and the exfoliated sample remained stable for the entire monitoring period (72 and 32 days, respectively) under ambient conditions. We further investigated UV laser treatment in a controlled-environment chamber under argon, nitrogen, and oxygen atmospheres, distinguishing the influence of oxygen as the PL treatment agent. These findings establish a reliable pathway for the permanent treatment of single-layer $MoS_2$ PL properties, an essential step toward practical, high-performance nanophotonic devices.

## INTRODUCTION

Single-layer transition metal dichalcogenides (TMDCs), with their unique optical properties and electronic structures, are promising for two-dimensional (2D) optoelectronic applications, such as photodetectors, photovoltaic devices, and light emitters.[1] TMDCs with the chemical formula $MX_2$ (M = Mo, W; X = S, Se), such as $MoS_2$, $WS_2$, $WSe_2$, and $MoSe_2$, exhibit an indirect-to-direct bandgap transition with a large exciton binding energy ranging from 1.5 to 2.0 eV when thin down to a monolayer. Therefore, their photoluminescence (PL) quality is stronger than that of their bulk counterparts. Consequently, excitons dominate the optical and electrical properties of these materials.[2–6] Unlike bulk materials, excitons in 2D semiconductor materials dictate the optical properties even at room temperature.[7] Most 2D TMDCs show a direct bandgap with a large excitonic binding energy in the visible wavelength range.[8] This binding energy can be modified by an external electric field, and thus the exciton resonances can be tuned or even switched off.[7] Generally, the light interaction with atomically thin materials such as TMDCs is weak. On top of that, even though 2D TMDCs should show higher PL ability in contrast to bulk counterparts, in reality, they exhibit a poor photoluminescence quantum yield (PLQY) (the ratio of the number of emitted photons to the number of absorbed photons) due to the presence of defects.[9] Chalcogen (S and Se) vacancies are the most common defects observed in both TMDCs monolayers exfoliated from bulk crystals and synthesized by chemical vapor deposition (CVD). Therefore, "healing" the defects of TMDCs could bring a significant improvement in the optical properties of TMDCs, as claimed in[7] and already proved in.[10]

The single-layer $MoS_2$ is non-centrosymmetric and has a direct bandgap that leads to high current on-off ratios, high optical absorption around 670 nm, strong PL, high refractive index, and efficient valley and spin control, which makes it suitable for photonic and optoelectronic nano-scale applications.[11,12] The optical properties of single-layer $MoS_2$ are dominated by the $A^o$ exciton (~1.88 eV) and many-body excitons, i.e., $A^-$ trion (~1.84 eV), and B exciton (~2.02 eV), which are prominent excitonic features.[11,13–15] Despite these strong light-matter interactions, the practical application of the single-layer $MoS_2$ in optoelectronic devices is severely hindered by its exceptionally low QY, often reported to be less than 1%.[16,17] This poor efficiency is primarily attributed to a high density of defect states.[18] Furthermore, the exciton population in $MoS_2$ can be divided into propagating excitons, which couple strongly with photons, and localized excitons, which become tightly bound to these defect sites, thereby quenching the PL.[11,12]

Intrinsic lattice defects in $MoS_2$, such as vacancies, substitutional impurities, adatoms, antisite defects, and grain boundaries, as byproducts of the fabrication process, play a critical role in modulating PL intensity and QY.[18–20] Sulfur vacancies are the most probable defects in mechanically exfoliated and CVD-grown $MoS_2$ single layers, which create a mid-gap state in the band structure that causes nonradiative recombination sites by trapping electrons,[2,5] and leads to n-doped single layers due to the unpaired electrons in sulfur vacancies, which affect the optical properties, where charge carriers lose energy without emitting photons.[21] Also, the electron donor defects help form trions $A^-$ by providing electrons to bind to excitons. Therefore, healing the defects increases the PL and QY. Various procedures, such as chemical treatment,[22] plasma treatment,[23] and laser irradiation[2,5,16,24] have been implemented recently for surface passivation of $MoS_2$ single layers to enhance the PL quality. Ambient atmospheric oxygen binding to Mo atoms at defect sites depletes the negative charges of n-doped $MoS_2$ and facilitates defect passivation.[6,25]

As for the laser treatment, numerous applications of visible laser radiation have been reported until now. The previously reported laser-based techniques suffer from several practical drawbacks that limit their widespread application. For instance, some methods rely on aggressive high-temperature annealing (300-500 °C) in vacuum or $H_2$ environments as a critical activation step to achieve PL brightening.[2,6,25] Such thermal pretreatments complicate the fabrication process and may induce structural damage, such as the formation of physical cracks.[25] Additionally, some techniques often result in non-homogeneous PL enhancement, spatially confined to localized defect sites rather than being uniform across the sample surface.[16,25] High laser power densities are often required to achieve significant enhancement. At these levels, the thermal effect of laser irradiation induces the evaporation of sulfur atoms, leading to permanent structural damage and a physical reduction in the material's thickness.[16] Also, chemical treatments using bis(trifluoromethane)sulfonimide (TFSI) superacid, which have demonstrated near-unity quantum yields in exfoliated flakes, remain largely ineffective for as-grown CVD monolayers, attributed to the inherent tensile strain induced by the growth substrate, which chemical "healing" of vacancies alone cannot resolve.[26] Furthermore, the resulting PL treatments are often characterized by limited temporal stability, with reports confirming persistence for only 25 to 72 hours under ambient or vacuum conditions.[2,16] In some cases, the PL enhancement is found to be entirely and quantitatively reversible, so that the signal immediately returns to its original value upon removal of the gas environment or exposure to vacuum.[6,16] This occurs because the modulation relies on the weak physical adsorption (physisorption) of molecules, which acts as a transient molecular gating force rather than providing permanent chemical passivation.[6] Additionally, many of these techniques are restricted to aged monolayers where the high density of chalcogen vacancies resulting from storage periods of up to 18 months is necessary to facilitate the observed brightening.[5,27] All these factors make the published laser techniques less suitable for achieving efficient, long-lasting improvements for the intrinsic optical properties of pristine, freshly prepared $MoS_2$ single layers.

In this paper, we report on the surface treatment of exfoliated and CVD $MoS_2$ single layers by ultraviolet (UV) laser radiation under ambient air conditions. This procedure sustainably improves the PL properties of these layers. It overcomes the limitations of previous laser treatments by processing a reasonable sample area and requiring no additional technological arrangements, such as a special environment. Hence, it represents a valuable routine method in the process of defect healing in single-layer $MoS_2$ materials and, possibly, other TMDC materials, leading to a durable increase in their PLQY and thus to new photonics applications

## RESULTS AND DISCUSSION

### PL Enhancement of $MoS_2$

An optical image of an exfoliated $MoS_2$ single-layer flake on a c-cut sapphire substrate treated by a continuous-wave (cw) UV laser (355 nm, 4 mW) is shown in Figure 1a. The sample was repeatedly scanned 10 times under atmospheric conditions over the laser beam, focused via a 40x objective, with a 200 nm pitch between each two processed points and a 0.5 s integration time per point. Figures 1b and 1c depict the PL maps before and after the UV laser treatment provided by a cw green excitation laser (532 nm, 0.3 mW) and a 100x objective. The pitch and integration time for taking the spectra over the sample area were 200 nm and 0.1 s, respectively. According to the averaged PL spectra in Figure 1d, taken from the UV-treated sample areas, the UV laser treatment induced an 8.8-fold increase in the PL peak intensity (height). Additionally, the signal intensity corresponding to the area under the PL peak shows a 6-fold increase. The difference between the peak height and the integrated area under the PL peak is attributed to a substantial narrowing of the peak expressed by the reduction of the peak's Full Width at Half Maximum (FWHM) (see below), suggesting a substantial improvement in the QY of the $MoS_2$ single layer. A primary advantage of this UV laser-assisted enhancement is its dual functionality, simultaneously durably passivating sulfur vacancies and modulating the electronic state (carrier population) of the single layers. By 'healing' the mid-gap states through oxygen chemisorption (see Figure 3 and Figure 5), the treatment effectively eliminates non-radiative recombination centers. This structural healing is complemented by a concurrent electronic shift, in which oxygen bonding induces p-type doping, thereby reducing excess electron density. This transition effectively converts radiatively inefficient trions into bright neutral excitons. Hence, such a treatment increases the total number of available radiative excitons, thereby maximizing PL of the $MoS_2$ single layers. To analyze the changes in the spectra, the PL spectra before and after the UV laser treatment were normalized, as shown in Figure 1e. The pristine sample exhibits a relatively broad emission peak (FWHM = 0.09 eV) with a prominent low-energy shoulder, characteristic of a significant trion contribution. After the UV laser treatment, the PL spectrum becomes considerably narrower (FWHM = 0.06 eV), and the low-energy shoulder is almost suppressed. This narrowing of the FWHM and the shift toward higher energies strongly indicate a change in the dominant excitonic species, a suppression of trions, and an enhancement of the neutral exciton emission.

To further quantify this change, the PL spectra were deconvoluted into their primary excitonic components: the neutral exciton ($A^o$), the negatively charged trion ($A^-$), and B-exciton, as seen in Figures 1f, g.

Figure 1f shows the deconvolution for the pristine sample. The spectrum is clearly trion-dominated, with the peak centered at ~1.86 eV, while the neutral exciton at ~1.89 eV appears as a smaller, higher-energy shoulder.[28] This behavior is typical for $MoS_2$ single layers, where native defects (such as sulfur vacancies) provide excess electrons, leading to n-type doping and the preferential formation of trions.[25] In contrast, Figure 1g shows the spectrum after the PL enhancement process. The deconvolution reveals a dramatic reversal in the exciton-trion ratio. The spectrum is now overwhelmingly dominated by the neutral exciton at ~1.88 eV. The trion peak at ~1.86 eV is significantly reduced to a minor component of the total emission. This transition from a trion-dominant to an exciton-dominant regime is the key mechanism behind the observed PL enhancement. By effectively reducing the defect-induced electrons, our treatment process passivates the defects, inhibits the formation of trions, and promotes the creation of neutral excitons, which provide much higher PL. The peak maximum position remains nearly unchanged, reflecting a balance between the suppression of lower-energy trions[28] and a subtle red shift of the neutral exciton peak, probably caused by strain relaxation.[29]

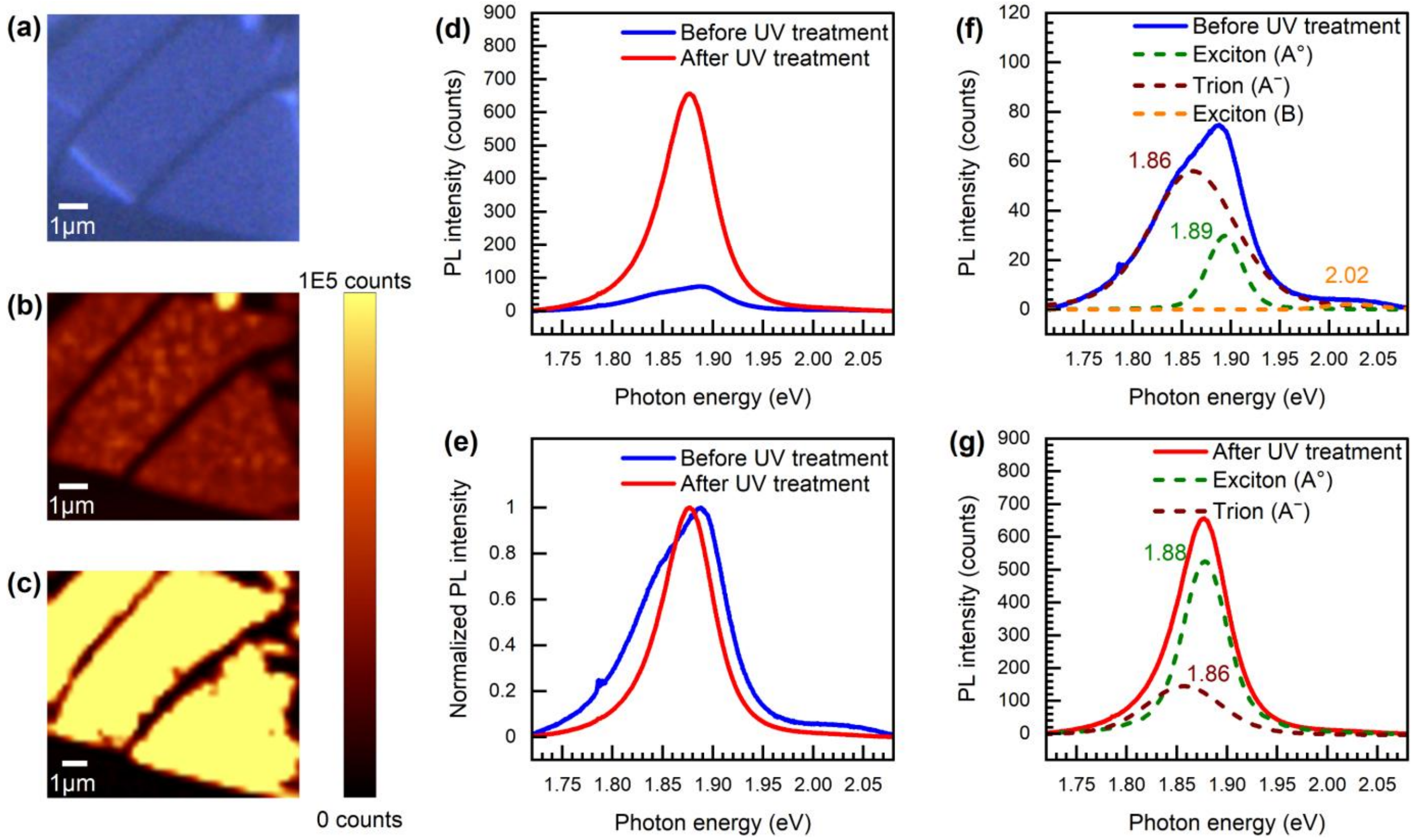

**Figure 1.** Enhancement of PL for an exfoliated $MoS_2$ single-layer flake treated by a UV laser ($\lambda_{exc}$ = 355 nm). (a) Optical image of the $MoS_2$ single-layer flake. (b) PL map obtained by the green excitation laser ($\lambda_{exc}$ = 532 nm) before UV laser treatment. (c) PL map obtained by the green excitation laser after UV laser treatment (the maps (b) and (c) have the same color scale). (d) The diagram shows an 8.8-fold increase in peak intensity (height) and a 6-fold increase in signal intensity (area under the peak) after UV laser treatment of the flake. (e) Normalized PL spectra before and after UV laser treatment, showing a significant spectral narrowing and the suppression of the low-energy trion shoulder, which indicates a transition to a more spectrally neutral exciton emission. (f) Fitted overall PL peak taken before UV laser treatment showing an exciton peak at 1.89 eV, a trion peak at 1.86 eV, and B-exciton at 2.02 eV. (g) Fitted overall PL peak after UV laser treatment showing the exciton peak at 1.88 eV and the trion peak at 1.86 eV.

To further validate the effectiveness and versatility of our enhancement method, we performed the same analysis on a CVD-grown $MoS_2$ single-layer sample. This allows us to test our process on a material with a possibly different defect nature and initial doping level. Similarly to the exfoliated sample, CVD-$MoS_2$ single-layer domains were treated with UV laser irradiation ($\lambda_{exc}$ = 355 nm) using a different set of parameters. The domain in Figure 2a was scanned at a laser power of 0.1 mW for 5 times, with a pitch of 200 nm between processed points and an integration time per point of 0.1 s. Figures 2b and 2c show the PL maps before and after the treatment, respectively, achieved using the same green laser (532 nm, 0.3 mW) and objective (100x) as in the case of the exfoliated $MoS_2$. After the treatment, the PL peak intensity (height) showed an 8.7-fold increase, and the signal intensity (area under the peak) increased 6.5-fold, as shown in Figure 2d.

Figure 2e presents the normalized PL spectra for a direct comparison of the spectral changes induced by the UV laser treatment. Similar to the exfoliated $MoS_2$, the pristine sample PL peak, showing up approximately at 1.84 eV, is relatively broad (FWHM = 0.08 eV). The spectrum of the treated sample shows two distinct changes: the peak maximum shifts to a higher energy of ~1.85 eV, and the overall linewidth, or FWHM, is reduced to 0.06 eV. Once again, this substantial spectral narrowing with the slight blue shift indicates a fundamental qualitative change in the excitonic population - a shift from trions to neutral excitons. To quantify this change, the PL spectra were deconvoluted into their constituent components - the neutral exciton and the negatively charged trion, as presented in Figures 2f and 2g.

Figure 2f shows the deconvolution of the pristine CVD $MoS_2$ peaks into three primary components: the neutral exciton ($A^o$) centered at 1.84 eV, the negatively charged trion ($A^-$) at 1.82 eV, and the B-exciton at 1.94 eV. In the pristine state, the emission is dominated by the trion component, which is more intense than the neutral exciton, indicating a high initial electron density. This trion-dominant state, combined with the presence of non-passivated sulfur vacancies, results in a relatively low overall QY. Finally, Figure 2g shows the PL peak fitting after the UV laser treatment. The spectrum is now overwhelmingly dominated by the neutral exciton at ~1.85 eV, while the trion contribution at 1.83 eV is significantly suppressed. This PL shift is similar to that observed for the exfoliated sample and confirms that our process effectively passivates sulfur vacancies via oxygen chemisorption (see Figure 3 and Figure 5). By blocking these non-radiative decay channels and simultaneously reducing the excess electron density (p-doping), the treatment significantly increases the radiative efficiency of the exciton population and redirects the recombination pathway to the more efficient neutral exciton state.

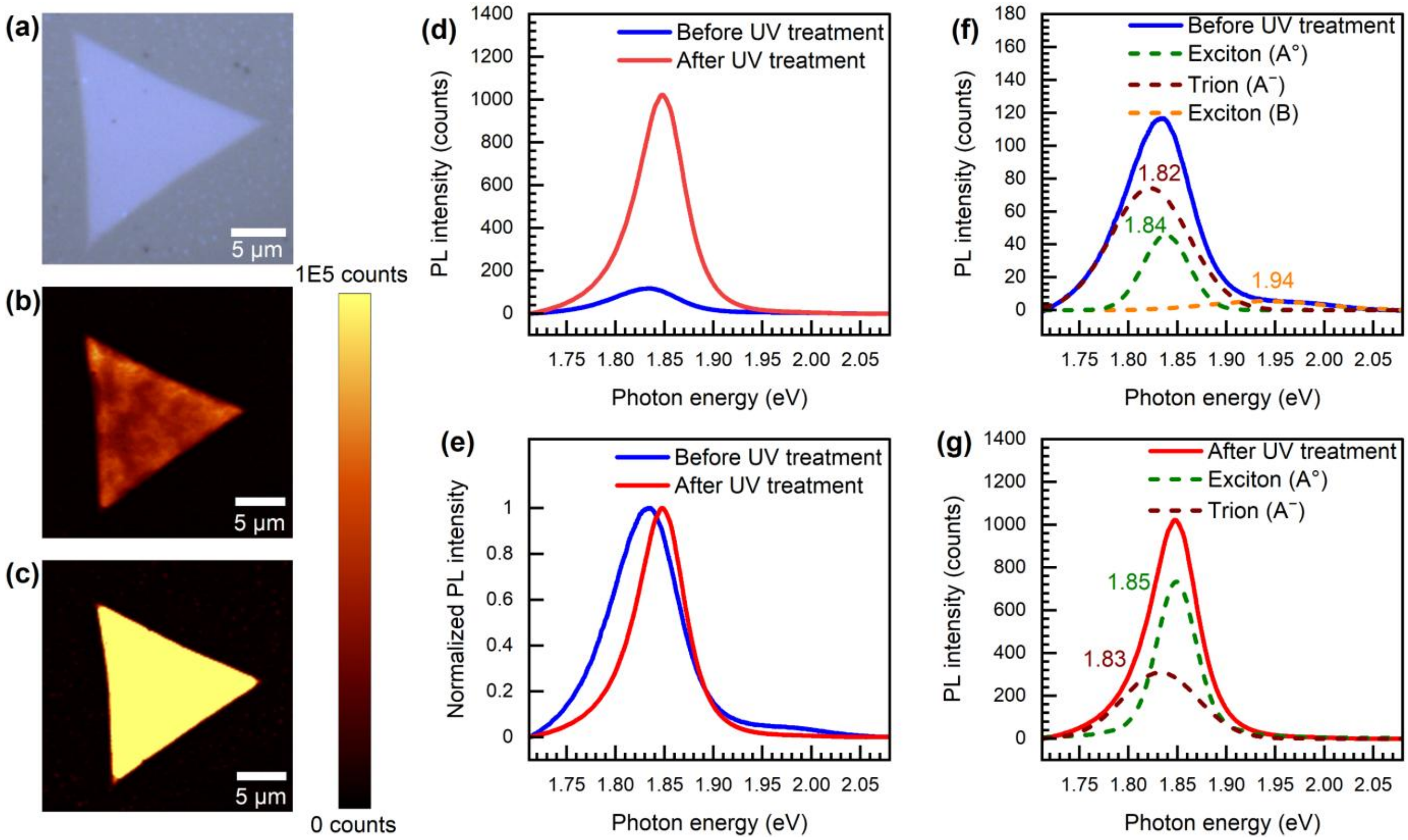

**Figure 2.** Enhancement of PL for a CVD $MoS_2$ single-layer domain treated by a UV laser ($\lambda_{exc}$= 355 nm). (a) Optical image of the domain. (b) PL map obtained by the green excitation laser ($\lambda_{exc}$ = 532 nm) before UV laser treatment. (c) PL map obtained by the green excitation laser after UV laser treatment (maps (b) and (c) have the same color scale). (d) The diagram shows an 8.7-fold increase in peak intensity (height) and a 6.5-fold increase in signal intensity (area under the peak) after UV laser treatment. (e) Normalized PL spectra before and after UV laser treatment showing a spectral narrowing and a distinct blue-shift of the peak maximum, which corresponds to the suppression of the low-energy trion ($A^-$) and the enhanced dominance of the neutral exciton ($A^o$) emission. (f) Fitted overall PL peak taken before UV laser treatment showing an exciton peak at 1.84 eV, a trion peak at 1.82 eV, and B-exciton at 1.94 eV. (g) Fitted overall PL peak taken after UV laser treatment showing the exciton peak at 1.85 eV and the trion peak at 1.83 eV.

## Raman Spectroscopy of $MoS_2$ Single-Layer

To investigate the physical mechanism of PL treatment and distinguish between mechanical strain relaxation and electronic doping effects, we performed Raman spectroscopy measurements before and after UV laser treatment (Figure 3). Pristine exfoliated sample exhibits the characteristic in-plane ($E^1_{2g}$) and out-of-plane ($A_{1g}$) vibrational modes centered at 384.7 $cm^{-1}$ and 404.4 $cm^{-1}$, respectively (Figure 3a). Also, the pristine CVD sample shows $E^1_{2g}$ mode at 382.9 $cm^{-1}$ and $A_{1g}$ mode at 403.9 $cm^{-1}$ (Figure 3b). After UV laser treatment, the position of the in-plane $E^1_{2g}$ mode for the exfoliated and CVD sample shifted to 385.2 $cm^{-1}$ and 383.6 $cm^{-1}$, respectively, representing a corresponding 0.5 $cm^{-1}$ and 0.7 $cm^{-1}$ blue shift. These shifts are directly attributed to the chemisorption of oxygen onto the single-layer $MoS_2$ surface sulfur vacancies. The formation of stable Mo-O bonds introduces a localized compressive strain within the single layer, which increases the frequency of in-plane vibrations.[30] Also, the $A_{1g}$ mode exhibits a distinct blue shift of roughly 0.7 $cm^{-1}$ for the exfoliated sample, moving to 405.1 $cm^{-1}$, and a 0.4 $cm^{-1}$ blue shift for the CVD sample, moving to 404.3 $cm^{-1}$. The $A_{1g}$ vibrational mode is highly sensitive to the charge-carrier concentration, with a shift to higher energy indicating a reduction in electron density (p-doping).[25]

Consequently, this Raman signature supports our proposed mechanism; the UV laser treatment effectively removes excess electrons and increases compressive strain via oxygen chemisorption. This depletion of the electron population suppresses the trion-formation probability and redirects combination pathways toward the more radiatively efficient neutral exciton state, thereby driving a significant enhancement in PL.

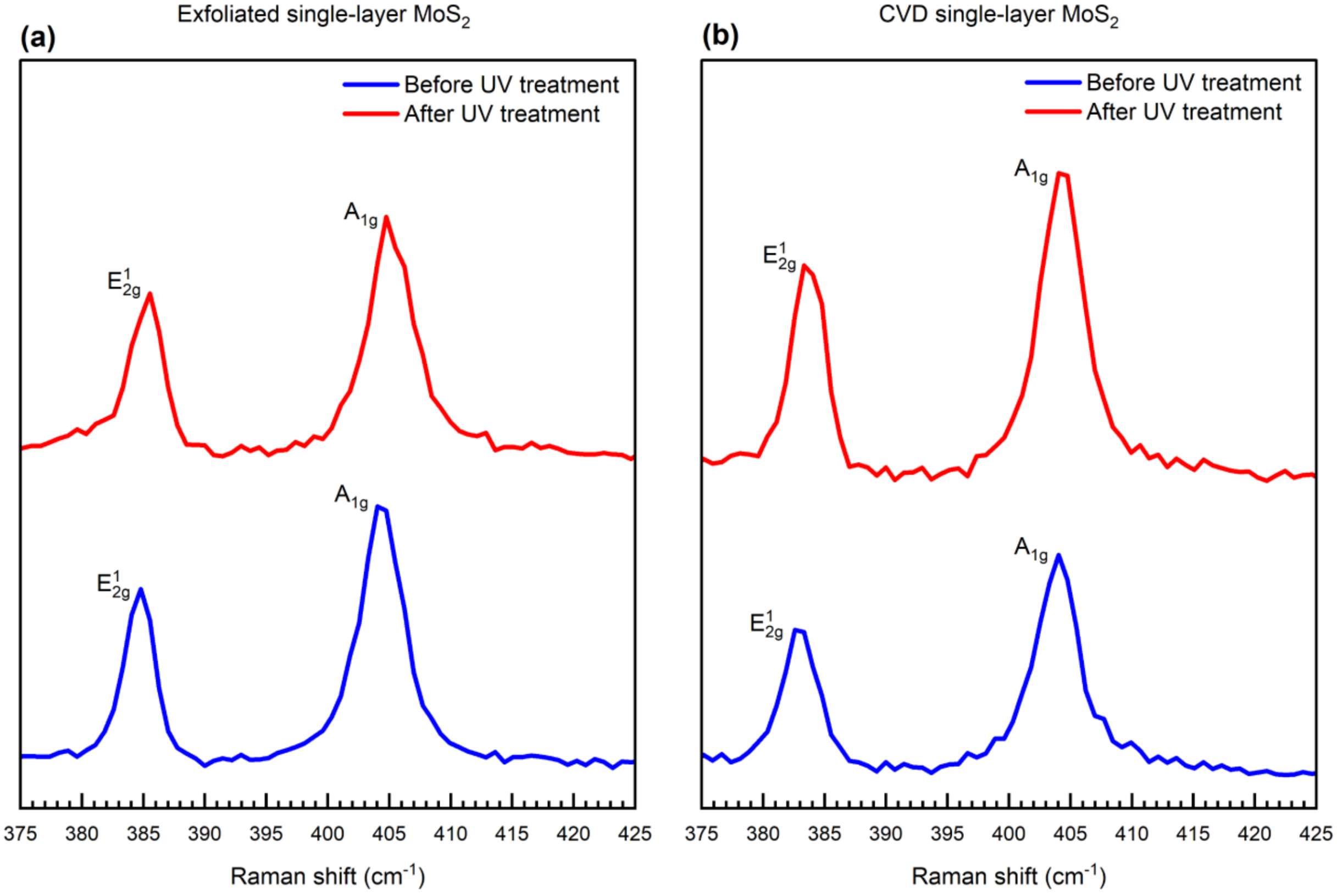

**Figure 3.** Raman spectra of the exfoliated $MoS_2$ single-layer and CVD $MoS_2$ single-layer domain before and after UV laser treatment. The spectra display the blue shift of the two primary vibrational modes, $E^1_{2g}$ and $A_{1g}$ for both exfoliated and CVD samples. (a) For the exfoliated sample, the in-plane $E^1_{2g}$ peak blue shifted by 0.5 $cm^{-1}$, indicating an increase in compressive strain due to the formation of the Mo-O bonds, and the out-of-plane $A_{1g}$ peak shows a 0.7 $cm^{-1}$

blue shift, attributed to a decrease in excess electron density (p-doping) resulting from UV laser-induced oxygen passivation. (b) For the CVD sample, the in-plane $E^1_{2g}$ peak position and the out-of-plane $A_{1g}$ peak blue shifted by 0.7 $cm^{-1}$ and 0.4 $cm^{-1}$ because of the Mo-O bonds formation and p-doping, respectively. The shifts have been determined via fitting the spectra with Lorentzian functions.

**Spatially Selected PL Treatment**

To demonstrate the high spatial precision of our laser processing method, we performed a localized enhancement experiment on a CVD-grown $MoS_2$ single layer. Figure 4a displays the optical image of a single-layer flake (domains), where a specific square region (outlined in purple) was targeted for the UV laser radiation.

The quantitative impact of the treatment is detailed in Figure 4b, which compares the averaged PL spectra extracted from the UV laser-irradiated region. The data reveal the same 5.9-fold increase for both the PL peak intensity (height) and signal intensity (area under the peak). This observation, together with the fact that FWHM remains remarkably constant at 0.05 eV, confirms that the spectral profile is preserved after the treatment (Figure 4b inset). Consequently, the selective PL treatment is primarily a result of structural healing - the passivation of the sulfur vacancies - rather than a net change in charge carrier concentration (p-doping), which would result in a trion reduction (and thus the peak shape change).

The difference in doping behavior between the selective-area and the whole CVD single-layer under UV laser treatment (Figure 2) is likely due to the lateral charge compensation from the untreated regions. When only the center is irradiated, the treated area is surrounded by a relatively large untreated area of n-type $MoS_2$ that acts as a continuous electron reservoir, which laterally supplies electrons to the treated zone.[31] This prevents the electron depletion required for a trion-to-exciton transition and prevents the local Fermi level from dropping into the neutral-exciton regime.

Before the treatment, the PL intensity map of the area (Figure 4c) shows a uniform but relatively weak signal, typical of the as-grown material, which contains a high density of non-radiative recombination centers. However, after localized laser scanning, the targeted region undergoes a significant modification. As visualized in Figure 4d, the radiated area exhibits intense PL, contrasting sharply with the surrounding pristine material. The treatment is spatially confined to the targeted geometry. The edges of the patterned region are distinct, with a minor intensity gradient at the boundaries attributed to the Gaussian intensity profile of the laser beam.

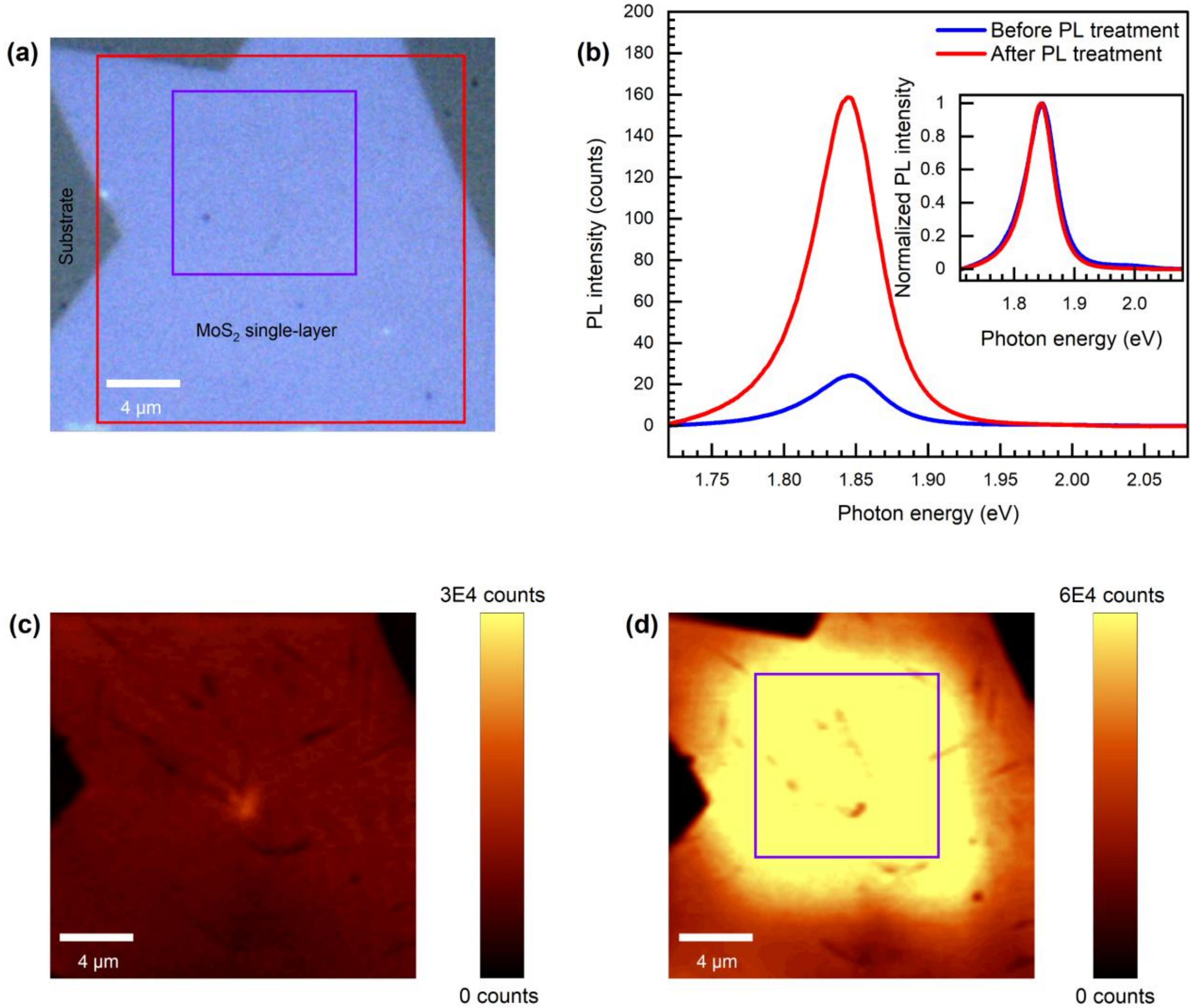

**Figure 4.** Spatially selected PL treatment of a CVD $MoS_2$ single-layer domain. (a) Optical image of the domain: the purple square corresponds to the UV-laser-illuminated area, and the red square to an approximate area for taking the PL map with a green laser ($\lambda_{exc}$ = 532 nm). (b) The diagram shows a 5.9-fold increase in PL intensity (peak height) in the marked area (the inset shows normalized PL spectra). (c) PL map taken by the green laser before UV laser treatment. (d) PL map taken by the green laser after UV laser treatment.

**Influence of Ambient Atmosphere on UV Laser Treatment**

To clarify the mechanism of UV laser treatment and verify the specific role of environmental oxygen, we performed UV laser treatment of a CVD $MoS_2$ single-layer domain under three distinct controlled gas atmospheres: argon, nitrogen, and oxygen.

The experiments were conducted in a custom-built chamber (Figure S3, Supplementary Information). In-situ PL measurements were carried out through a chamber window (c-cut sapphire) using a 40x objective (Figure 5), while ex-situ PL measurements were conducted immediately after taking the samples out of the chamber, using both 40x and 100x objectives (Figure S1, Supplementary Information). Figure 5 shows the evolution of the UV laser treatment inside the chamber. As demonstrated in Figures 5a and 5b, UV laser treatment in argon and nitrogen atmospheres resulted in a deterioration of the PL properties. This reduction indicates that in the absence of oxygen, the laser beam increases the number of non-passivated vacancies. Conversely, UV laser treatment under an oxygen atmosphere (Figure 5c) yielded a substantial increase in PL, reproducing the treatment observed in ambient air conditions. Furthermore, we performed a follow-up measurement immediately after taking the samples out of the chamber. We observed that upon re-exposure to the air, the samples that had been treated in inert gases (Ar, $N_2$) and exhibited PL quenching inside the chamber showed a spontaneous increase in PL intensity (Figure S1a, b, and S1c, d, respectively), without any further UV laser treatment. Also, the sample treated in $O_2$ retained its emission intensity outside the chamber (Figure S1e, f). This reversible behavior indicates that the mechanism is driven by dynamic oxygen adsorption to sulfur vacancies. Consequently, the PL quenching in the inert gas was actually caused by the loss of passivating oxygen, which was readily restored upon returning the sample to an air environment.

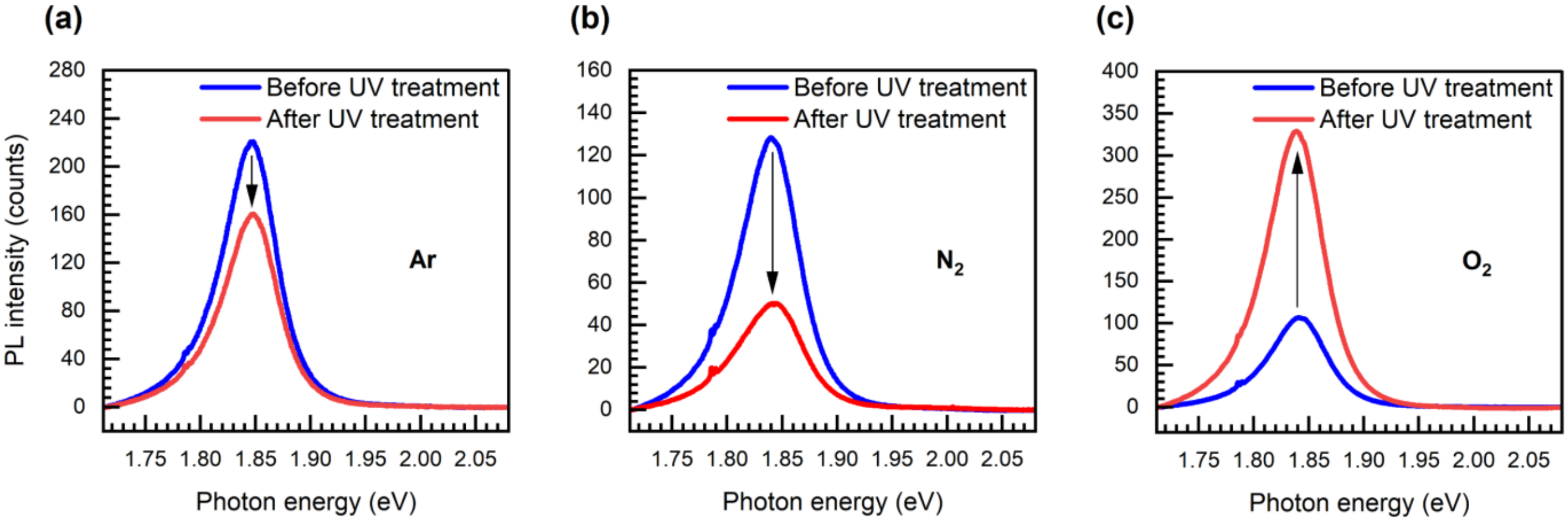

**Figure 5.** Impact of Ar, $N_2$, and $O_2$ gas atmosphere on the performance of UV laser treatment of a CVD $MoS_2$ single-layer domain. (a) In an Ar atmosphere, the PL intensity decreased after UV laser treatment. (b) In an $N_2$ atmosphere, a distinct quenching of the PL signal is observed following the UV laser treatment. (c) In an $O_2$ atmosphere, the PL intensity is significantly increased after UV laser treatment, confirming that oxygen adsorption to sulfur vacancies is the responsible mechanism for the UV laser treatment.

## Long-term Stability and Evolution of PL Treatment

A crucial aspect for any enhancement technique is its long-term stability and robustness under ambient conditions. To evaluate this, we monitored the time-dependent PL enhancement factor of the UV laser-treated samples, both exfoliated and CVD, over 32 and 72 days, respectively.

Figure 6a tracks the stability of the exfoliated single-layer over a 32-day interval. Immediately after UV laser treatment, the sample showed a sharp ~8-fold increase in PL intensity (peak height). While a distinct decrease (~30 %) was observed during the initial week, the signal subsequently reached values fluctuating within a 10 % interval. This behavior confirms that, despite the initial transient PL decrease, the treatment overall provides a significant and permanent PL improvement, indicating irreversible healing of structural defects and the treated monolayer's resistance to further ambient degradation.

Figure 6b illustrates the temporal evolution of the PL intensity for a CVD-grown $MoS_2$ domain after UV laser treatment. After UV laser treatment, the PL intensity (peak height) increased 4 times. After fluctuating within a 25 percent interval during the first 20 days, it maintained stable, elevated values showing no systematic decrease over time. Hence, once again, the graph demonstrates the stability and robustness of the UV laser treatment under air conditions.

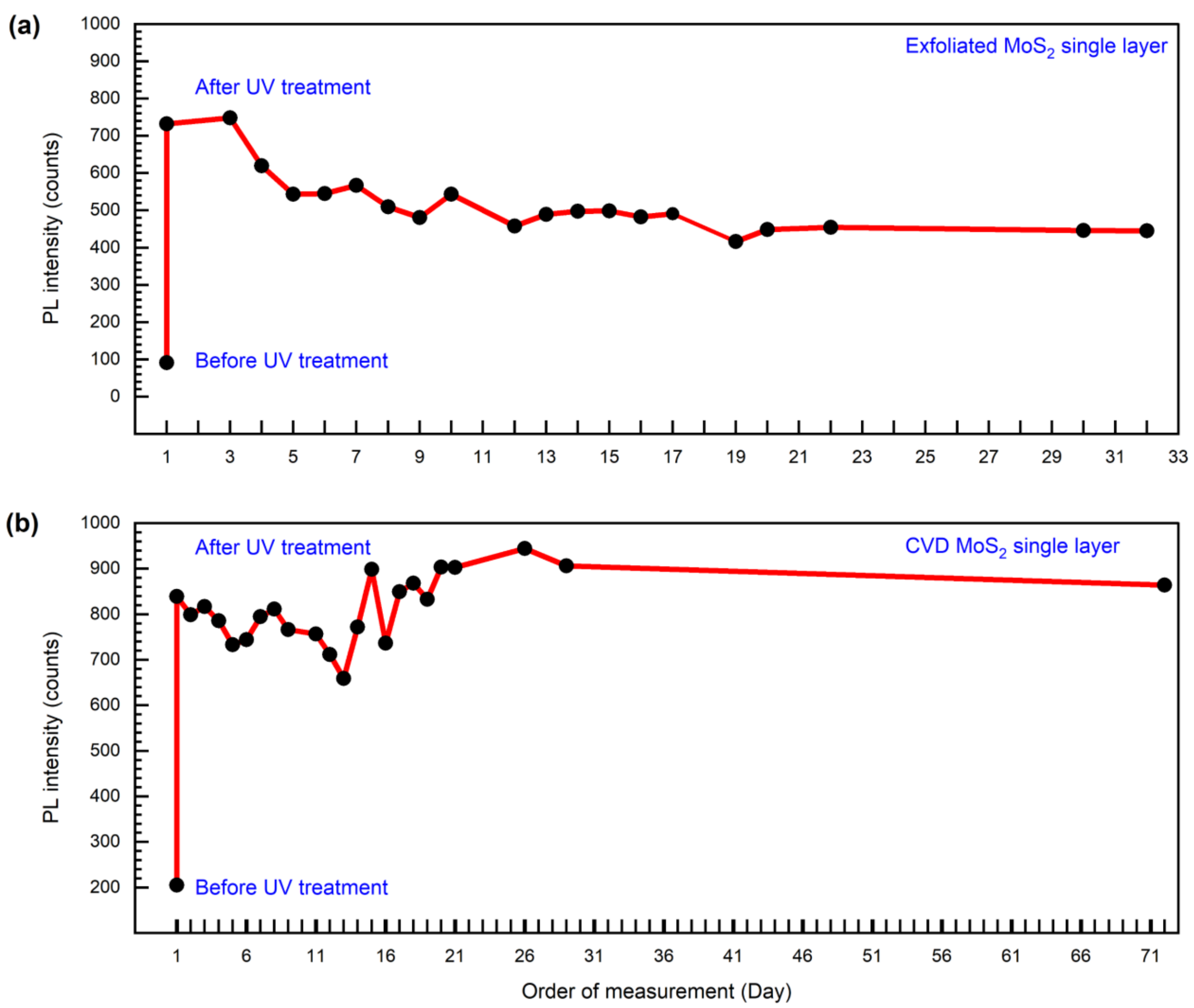

**Figure 6.** Long-term stability of the PL intensity for an exfoliated $MoS_2$ single-layer flake and a CVD $MoS_2$ single-layer domain after UV laser treatment. (a) The measurement on Day 1 shows an 8-fold PL increase (peak height) of an exfoliated $MoS_2$ single layer, followed by a 30% decrease in the PL signal over the next few days. In the rest of the days, the signal fluctuates within a 10 percent interval around a constant value. (b) The measurement on Day 1 shows the 4-fold PL increase (peak height) in the case of a CVD single-layer domain. After 20 days of signal fluctuations (25 %), the signal remains high and constant, indicating no observable degradation and confirming the permanence and stability of the enhancement.

## CONCLUSION

In conclusion, we have developed a robust, spatially precise method for a substantial PL enhancement of both exfoliated and CVD-grown $MoS_2$ single layers via controlled cw UV laser irradiation. We have observed a more than 8-fold maximal increase in the PL signal (in height), accompanied by a substantial spectral narrowing and a shift in the excitonic weight from trion to neutral-exciton-dominated emission. Detailed deconvolution of the PL spectra indicates that the treatment is driven by an effective passivation of sulfur vacancies, which serve as non-radiative recombination centers. This is confirmed by Raman spectroscopy, showing that the mechanism is purely electronic-driven by p-type doping, as evidenced by the distinct blue shift of the $A_{1g}$ mode and the blue shift of the strain-sensitive $E^1_{2g}$ mode due to the formation of Mo-O bonds.

Our controlled-atmosphere experiments within an environmental chamber clearly identify molecular oxygen as the critical agent in this process; while UV laser treatment in Ar and $N_2$ atmospheres leads to PL quenching, the presence of $O_2$ effectively facilitates the treatment of the PL signal. Furthermore, we have observed a significant reversible behavior of PL quenched in inert gas environments, showing a spontaneous recovery upon re-exposure to ambient air without further irradiation. This reveals that the laser's primary role is to activate surface sites, which subsequently bind atmospheric oxygen for radiative stabilization.

The demonstrated spatial precision for localized UV laser treatment, alongside the long-term stability exceeding 72 days for CVD domains and 32 days for exfoliated flakes, establishes this laser-assisted approach as a highly effective, chemical-free pathway for durable optimization of the optical performance of 2D semiconductors for next-generation nanophotonic devices.

## EXPERIMENTAL METHODS

Two types of single-layer $MoS_2$ samples were used for PL enhancement - small flakes and larger single-crystal domains. The $MoS_2$ single-layer flakes were prepared by mechanical exfoliation from a bulk crystal using a blue adhesive tape. First, we pressed a bulk $MoS_2$ crystal onto the tape, then folded the tape multiple times to cover the entire area with thin layers (single-layer and a few layers) of $MoS_2$. Then, the layers were transferred to polydimethylsiloxane (PDMS) blocks (Gel-Pak Company) on a hot plate (60 °C) for a better yield of the single-layer coverage. In the last step, the PDMS blocks were inverted and pressed face-down onto a double-sided-polished c-cut sapphire (thickness 430 ± 0.1 μm) at 90 °C to transfer the $MoS_2$ layers onto the substrate. The typical linear size of the single-layer flakes was up to 10 μm.

The larger, homogeneous $MoS_2$ single-layer domains were prepared by CVD on a double-sided-polished c-cut sapphire substrate (thickness 430 ± 0.1 μm). At the beginning, the substrate was cleaned by subsequent sonication in acetone and isopropanol (Sigma-Aldrich). Then, the $MoS_2$ layer was grown in a quartz tube of a horizontal furnace with two heating zones, under atmospheric pressure, using argon as the carrier gas. Argon was introduced into the tube via a gas line attached to one end, which then led to a bubbler filled with 100 mM aqueous KOH solution. The tube was flushed with argon at a continuous flow of 200 $cm^3\ min^{-1}$ at a temperature of ~25 °C for 15 minutes. The precursor, $MoO_2$ (30 mg) powder (Sigma-Aldrich, LOT: STBH3472, purity 99%), was placed in the crucible in the first heating zone (high temperature), and the substrate was placed face down on top of the crucible. Simultaneously, 100 mg of sulfur (Thermo Scientific, LOT: T29G013, purity 99.9995%) was placed in a tube 20 cm away in the second heating zone (low temperature). Maintaining a constant argon flow rate of 120 $cm^3\ min^{-1}$, the temperature was subsequently increased in the first heating zone at a constant rate of 40 °C $min^{-1}$. When the temperature reached

~670 °C, the heating in the second zone was turned on. Finally, the temperature was held constant at ~820 °C in the first zone and ~180 °C in the second zone for 8 minutes. Afterward, the furnace was opened, and the system was cooled down to ambient temperature.

Irradiation of the $MoS_2$ single-layer structures (both exfoliated and CVD-grown) was performed using a cw UV laser ($\lambda = 355$ nm) in a Raman/PL setup (Alpha 300 R, WITec) at room temperature and under ambient air conditions to improve their optical quality. The influence of various parameters, including laser power, irradiation time, and the number of irradiation spots (i.e., the pitch between them), was explored to increase PL in $MoS_2$ single-layer objects without causing any destructive effects. The PL properties and Raman spectra of these layers were studied using the above-mentioned μRaman/PL system with a 532 nm laser (cw, WITec). To learn more about the mechanism of PL property improvement of $MoS_2$ layers, complementary experiments under oxygen, argon, and nitrogen atmospheres were carried out. The thickness and topography of the flakes were determined by atomic force microscopy (AFM) using a scanning probe microscope Dimension Icon (BRUKER).

## ACKNOWLEDGMENTS

This project has been supported by the project Quantum Materials for Applications in Sustainable Technologies (QM4ST), funded as project No. CZ.02.01.01/00/22_008/0004572 by OP JAK, call Excellent Research. The authors also gratefully acknowledge support from the CzechNanoLab project (ID 90251), funded by the Ministry of Education, Youth and Sports of the Czech Republic, for measurements carried out at the CEITEC Nano Research Infrastructure. We acknowledge CF Nanobiotechnology of CIISB, Instruct-CZ Centre, supported by MEYS CR grants no LM2023042 and LUC24105, and European Regional Development Fund-Project „Innovation of Czech Infrastructure for Integrative Structural Biology“ (No. CZ.02.01.01/00/23_015/0008175). Also, the JAROSLAV KOČA BRIDGE FUND 2024 project is acknowledged for its financial support.

## DATA AVAILABILITY

The data underlying this study are openly available at https://doi.org/10.5281/zenodo.20159866.

# Supporting Information for

# Durable Enhancement of $MoS_2$ Single-Layer Photoluminescence by Ultraviolet Laser Treatment Under Ambient Conditions

*Mahan Bakhshikhah,[1] Jiří Liška,[1,2] Rahul Kesarwani,[3] Jindřich Mach,[1,2] Ondřej Červinka,[1] Petr Dub,[1,2] Jiří Spousta,[1,2] Jan Přibyl,[4] Jana Kalbáčová Vejpravová,[3] and Tomáš Šikola[1,2*]*

[1]Central European Institute of Technology, Brno University of Technology, 612 00 Brno, Czech Republic

[2]Institute of Physical Engineering, Faculty of Mechanical Engineering, Brno University of Technology, 616 69 Brno, Czech Republic

[3]Department of Condensed Matter Physics, Faculty of Mathematics and Physics, Charles University, 121 16 Prague 2, Czech Republic

[4]Central European Institute of Technology, Masaryk University, 625 00 Brno, Czech Republic

[*]sikola@fme.vutbr.cz

**Reversibility of UV Laser-Induced PL Quenching in Ambient Air**

To further confirm the oxygen passivation mechanism, follow-up ex-situ PL measurements were performed on the laser-treated CVD $MoS_2$ single-layer domains immediately after removing them from the environmental chamber and placing them in ambient air. The measurements were initially conducted with a 40x objective to maintain consistency with the in-situ measurements performed inside the chamber. To ensure high spectral resolution and accurately monitor PL shifting, spectra were additionally acquired with a 100x objective, as illustrated in Figure S1.

For the sample treated in an Ar atmosphere (Figures S1a and S1b), the PL intensity, which had been suppressed during in-situ measurement, showed a significant spontaneous increase upon exposure to ambient air. Regardless of the objective used, the final emission level clearly exceeded the pristine PL. A similar recovery behavior was observed for the nitrogen-treated sample (Figures S1c and S1d), where the previously quenched emission recovered and subsequently exceeded the original intensity once the domain was re-exposed to the ambient air. The fact that the enhancement occurs only upon the introduction of air, despite the laser processing being completed in an inert environment, clearly identifies molecular oxygen as the external agent responsible for the observed UV laser treatment.

Finally, the sample treated in $O_2$ (Figures S1e and S1f) retained its enhancement both inside and outside the chamber. This consistent behavior confirms that the PL quenching observed in inert gases is a reversible state driven by the absence of a passivating oxygen agent rather than permanent structural damage.

This spontaneous recovery without further laser irradiation suggests that the UV treatment in inert gas environments successfully creates the necessary surface-active sites. However, in the absence of oxygen, these sites act as non-radiative recombination centers, quenching the PL. Once exposed to air, the rapid chemisorption of atmospheric oxygen molecules effectively passivates the defects on these laser-activated sites. These results demonstrate that oxygen is the fundamental requirement for PL enhancement; it acts as an electron-depleting agent, shifting the excitonic balance toward the more radiatively efficient neutral exciton state, thereby activating the latent enhancement mechanism within the inert gas environments. This atmospheric recovery provides a clear distinction between the laser's role in site activation and oxygen's role in PL radiation.

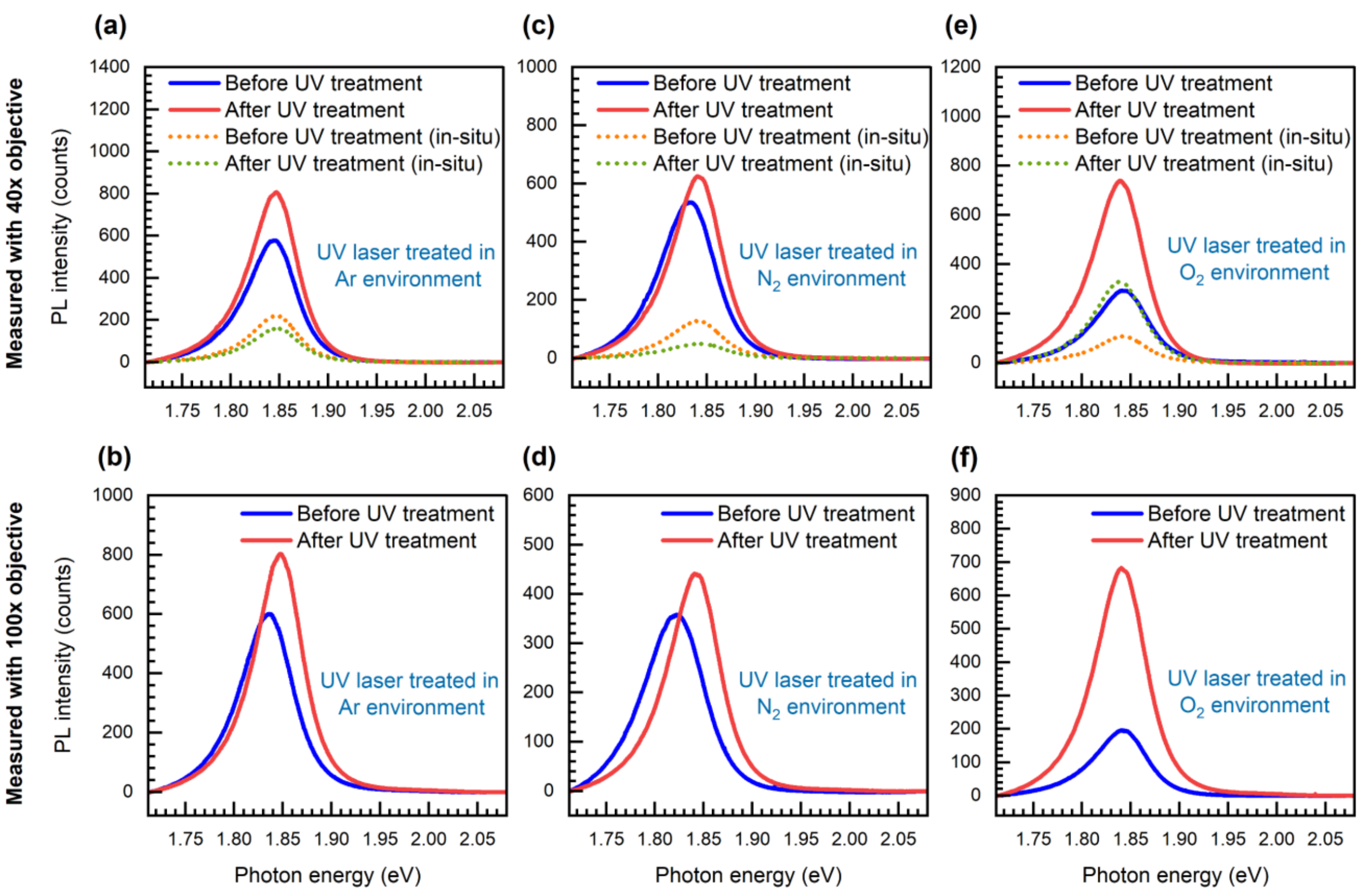

**Figure S1.** Ex-situ PL characterization in ambient air following the UV laser treatment in the Ar, $N_2$, and $O_2$ atmospheres. The spectra compare the emission before and after the treatment, measured with both 40x and 100x objectives under ambient air. For comparison, there are also the PL spectra measured in situ, conducted in the custom-built chamber just after the UV treatment (dashed curves). (a, b) Recovery and small enhancement of PL intensity in the ambient air for the sample previously quenched during treatment in an Ar environment. (c, d) Spontaneous PL enhancement was observed in air for the sample that showed a decreased emission during treatment under $N_2$. (e, f) Sustained high-intensity PL emission for the sample treated in an $O_2$ atmosphere, which retains its enhanced state upon exposure to ambient air.

**Nanoscale Topographical Characterization**

To ensure that the UV laser treatment is non-destructive and does not induce thermal artifacts or structural degradation, we performed AFM characterization before and after the treatment. As shown in Figure S2a, the surface morphology of the exfoliated $MoS_2$ single-layer, including characteristic features such as intrinsic cracks and small surface particles, remained entirely unchanged after treatment. This topographical stability was quantitatively confirmed by Root Mean Square (RMS) roughness measurements, which shifted from 0.648 nm to 0.522 nm after treatment. This slight reduction in roughness suggests a potential surface-cleaning effect, likely due to laser-induced desorption of contaminants or moisture rather than lattice degradation. Similarly, the CVD-grown $MoS_2$ single-layer domain exhibited remarkable physical robustness (Figure S2b). The RMS roughness showed negligible variation, moving from 1.580 nm to 1.571 nm, further confirming that the UV laser irradiation does not cause topographical alterations or material ablation.

These results demonstrate that the observed PL treatment is due to the defect passivation and is achieved without degrading the structural integrity or physical stability of the $MoS_2$ monolayers.

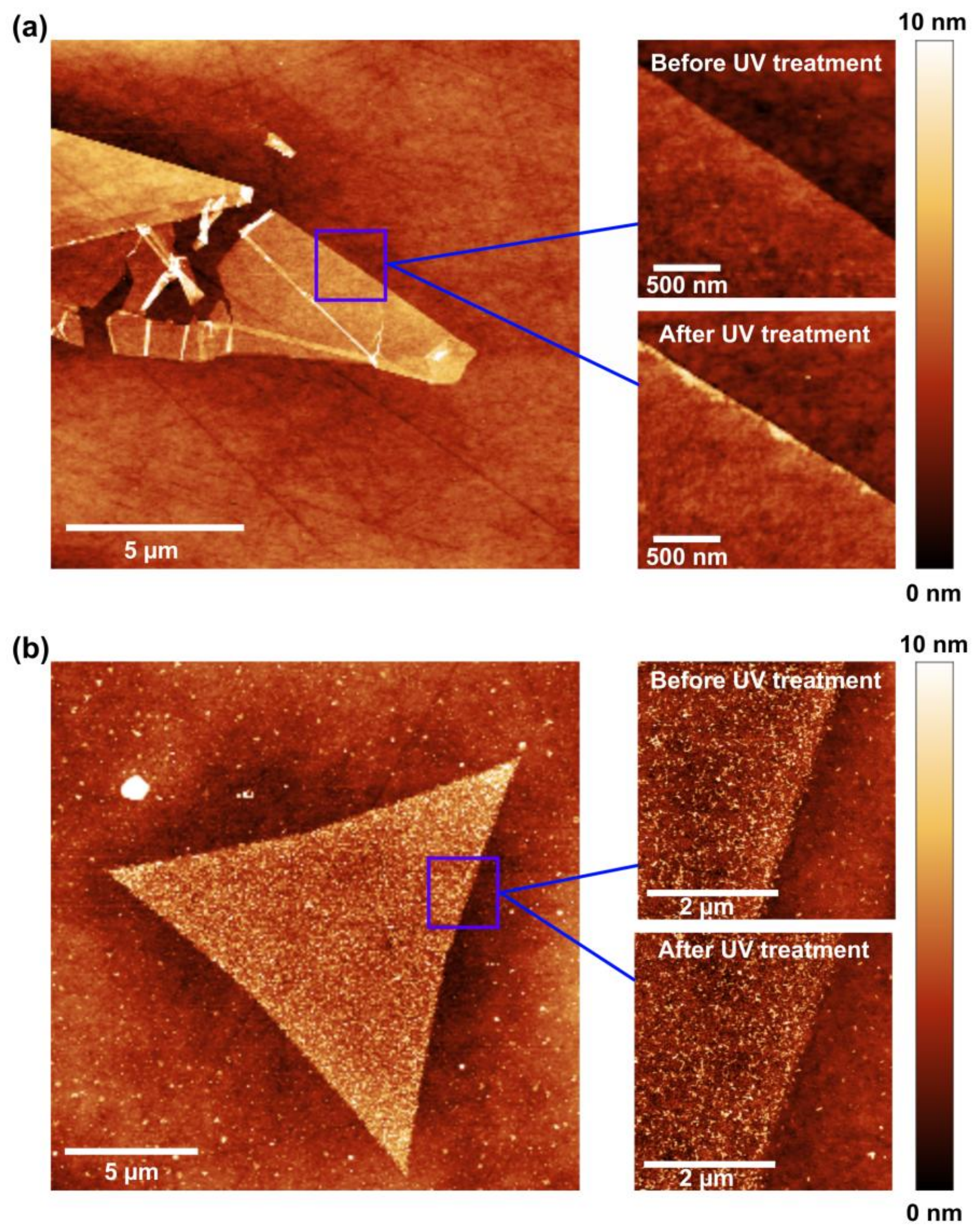

**Figure S2.** Comparative AFM height maps illustrate the surface morphology and topographical stability of (a) exfoliated and (b) CVD-grown $MoS_2$ single-layers recorded before and after UV laser treatment. The preservation of specific topographical features, including flake edges, intrinsic cracks, and surface particles, demonstrates the non-destructive nature of the PL treatment. Quantitatively, the RMS roughness of the exfoliated flake decreased from 0.648 nm to 0.522 nm, suggesting a surface-cleaning effect, while the CVD sample RMS roughness remained nearly constant, shifting from 1.580 nm to 1.571 nm. These results confirm that the PL sample treatment is left without lattice degradation or material ablation.

(a)

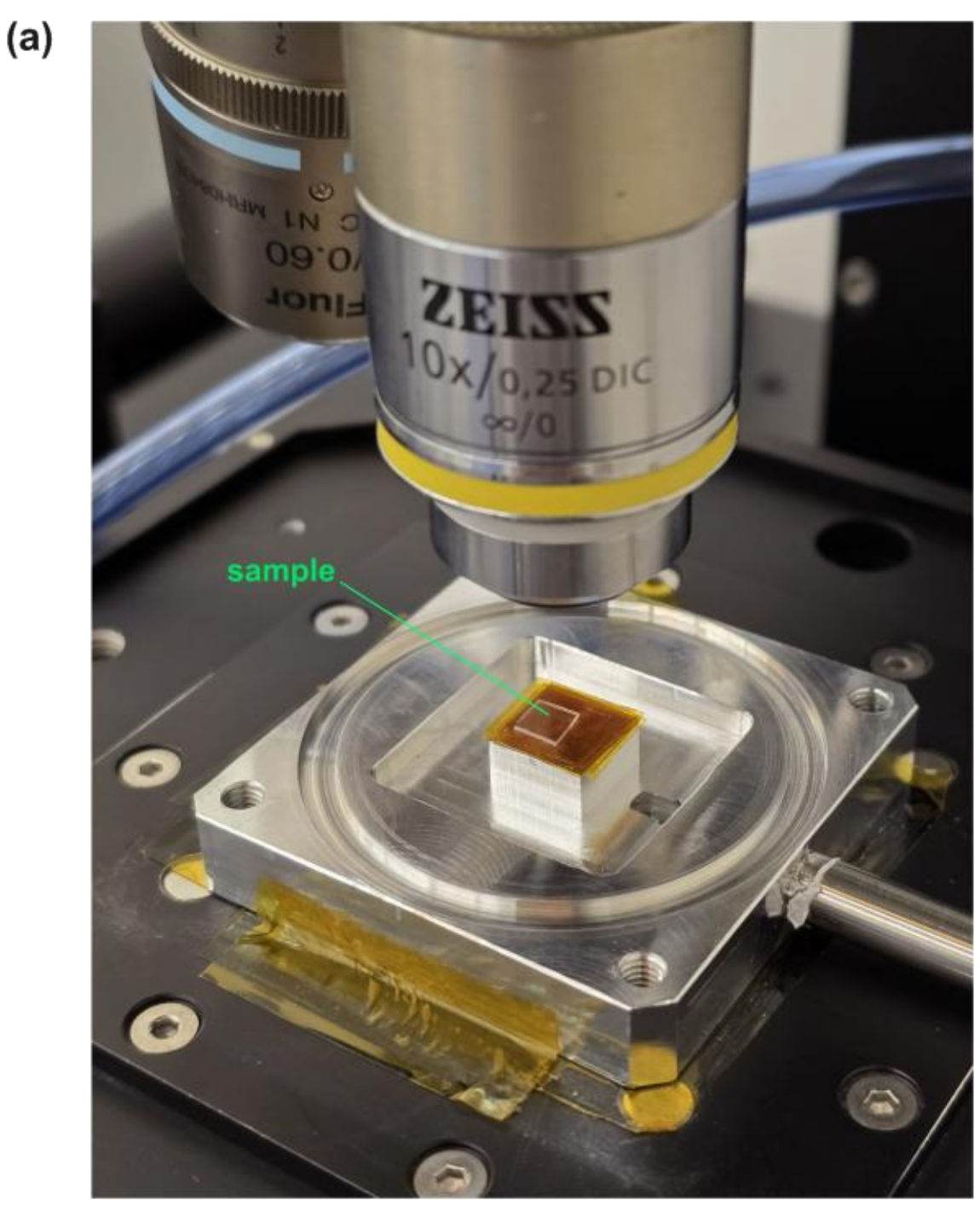

(b)

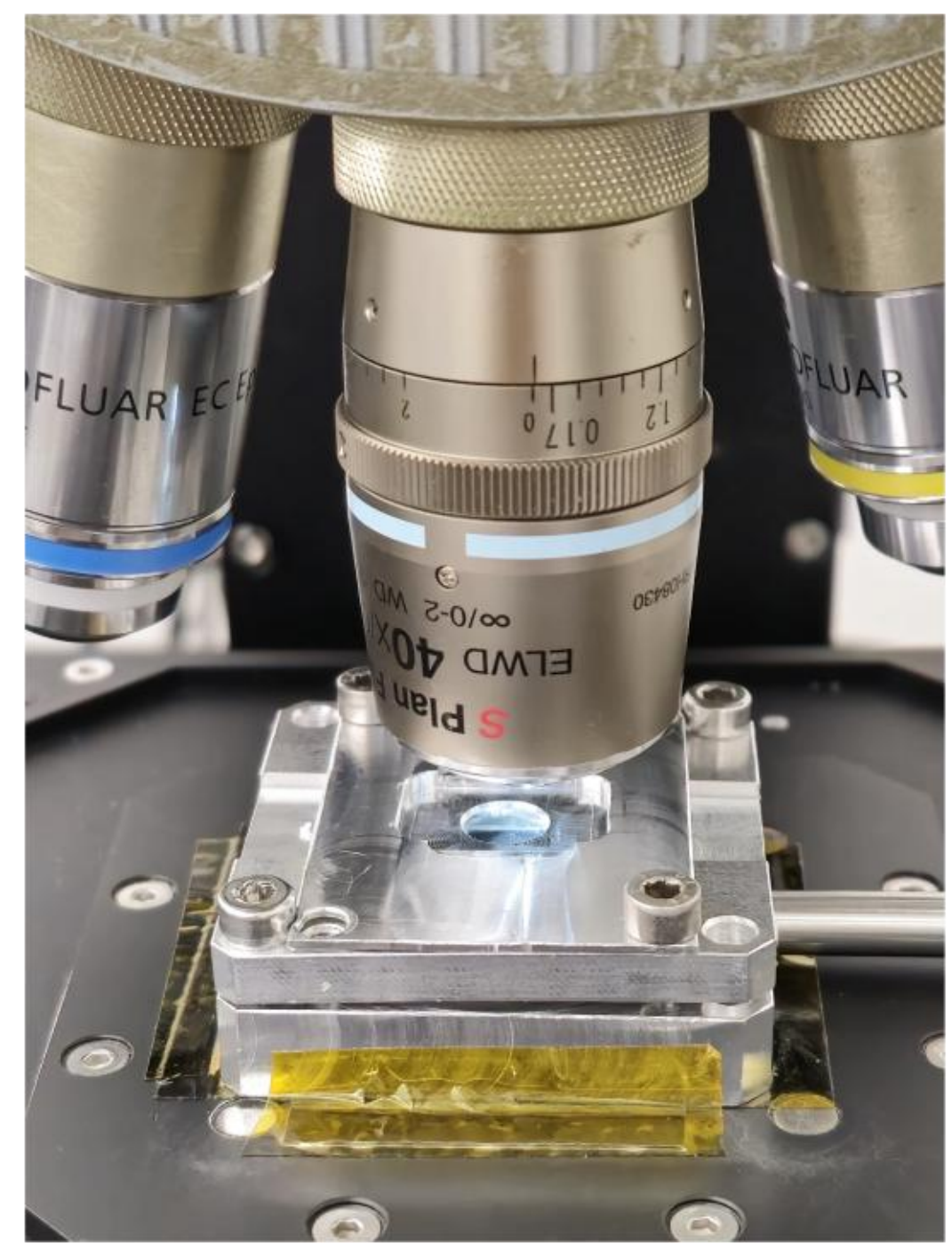

**Figure S3.** An experimental setup for UV laser treatment of $MoS_2$ under controlled Ar, $N_2$, and $O_2$ environments equipped with a gas inlet and a vacuum pump outlet. (a) Perspective view of the opened chamber revealing a sapphire substrate containing CVD $MoS_2$ single-layers (see the arrow). (b) The closed-chamber configuration for UV laser treatment provides a vacuum-tight environment of Ar, $N_2$, and $O_2$.